\documentclass[%
 reprint,
superscriptaddress,
showpacs,preprintnumbers,
 amsmath,amssymb,
 aps,
 pra,
]{revtex4-1}

\usepackage{graphicx}% Include figure files
\usepackage{dcolumn}% Align table columns on decimal point
\usepackage{bm}% bold math
\usepackage{amsmath,amsfonts,amssymb,amsthm}
\usepackage{txfonts}
\usepackage{enumerate}
\usepackage{color}
\usepackage{braket}

\makeatletter

\@addtoreset{thm}{section}
\makeatother

\makeatletter
\renewcommand{\theequation}{\arabic{section}.\arabic{equation}}
\@addtoreset{equation}{section}
\makeatother

\begin{document}

\preprint{APS/123-QED}

\title{Quasi-Nambu-Goldstone modes in nonrelativistic systems}
\author{Muneto Nitta}
\affiliation{Department of Physics, Keio University, Hiyoshi 4-1-1, Yokohama, Kanagawa 223-8521, Japan}
\affiliation{Research and Education Center for Natural Sciences, Keio University, Hiyoshi 4-1-1, Yokohama, Kanagawa 223-8521, Japan}
\author{Daisuke A. Takahashi}
\affiliation{Research and Education Center for Natural Sciences, Keio University, Hiyoshi 4-1-1, Yokohama, Kanagawa 223-8521, Japan}
\affiliation{RIKEN Center for Emergent Matter Science (CEMS), Wako, Saitama 351-0198, Japan}

\date{\today}% It is always \today, today,

\begin{abstract}
When a continuous symmetry is spontaneously 
broken in nonrelativistic systems, 
there appear either 
 type-I or type-II Nambu-Goldstone modes (NGMs) 
with linear or quadratic dispersion relation, respectively. 
When equation of motion or the potential term has 
an enhanced symmetry larger than that of Lagrangian or Hamiltonian, 
there can appear quasi-NGMs if it is spontaneously broken. 
 We construct a theory to count the numbers 
of type-I and type-II quasi-NGMs and NGMs, 
when the potential term has a symmetry of a non-compact group. We show that the counting rule based on the Watanabe-Brauner matrix is valid 
only in the absence of quasi-NGMs
because of non-hermitian generators, 
while that based on the Gram matrix [DT \& MN, arXiv:1404.7696, Ann. Phys. \textbf{354}, 101 (2015)] is still valid in the presence of 
quasi-NGMs. 
We show that there exist 
two types of type-II gapless modes, 
a genuine NGM generated by 
two conventional zero modes (ZMs) originated from the Lagrangian symmetry, 
and quasi-NGM generated by 
a coupling of one conventional ZM
and one quasi-ZM, which is originated from the enhanced symmetry, or two quasi-ZMs. 
We find that,  depending on the moduli, some NGMs can change to 
quasi-NGMs and vice versa
with preserving the total number of gapless modes.
The dispersion relations are systematically calculated by a perturbation theory.  
The general result is illustrated by the complex linear $O(N)$ model, 
containing the two types of type-II gapless modes 
and exhibiting the change between 
NGMs and quasi-NGMs.
\end{abstract}

\maketitle

\section{Introduction}
Symmetry principle is one of 
the most important concepts for modern physics. 
When a continuous symmetry of Hamiltonian or Lagrangian 
is not preserved in the ground state, 
spontaneous symmetry breaking (SSB) occurs 
\cite{Nambu:1961tp,Nambu:1961fr}.
SSB is ubiquitous in nature 
from magnetism, superfluidity and 
superconductivity to quantum field theories, 
in which it is the most important basis to achieve 
unification of fundamental forces. 
When such a SSB occurs,  
there must appear gapless modes 
known as Nambu-Goldstone modes (NGMs) 
\cite{Nambu:1961tp,Nambu:1961fr,Goldstone:1961eq}.  
NGMs are the most important 
degrees of freedom at low-energy 
\cite{Coleman:1969sm,Callan:1969sn,
Leutwyler:1993gf}.
In relativistic systems, 
 dispersion relations of NGMs 
are always linear.
On the other hand, 
the dispersion relation can be either linear $ (\epsilon \propto |k|) $  or quadratic  $ (\epsilon \propto k^2) $ 
in nonrelativistic systems. 
They are called type-I and type-II NGMs, respectively 
\cite{Nielsen:1975hm}.
Prime examples are given by 
the Heisenberg ferromagnets and antiferromagnets,
which give one type-II and two type-I NGMs, respectively,  
although symmetry breaking pattern is the same,
$SO(3) \to SO(2)$, and there are two broken 
generators for both cases.
Spinor Bose-Einstein condensates (BECs) 
of ultracold atoms 
\cite{Kawaguchi:2012ii,RevModPhys.85.1191}
provide a variety of examples 
of type-II NGMs \cite{DTMN}.  
In high energy physics, 
type-II NGMs appear in dense quark matter 
\cite{Miransky:2001tw,Schafer:2001bq,
Blaschke:2004cs}.

The number of NGMs coincides with the number 
of generators of broken symmetries 
in relativistic theories.
On the other hand, 
the number of NGMs in nonrelativistic systems 
has been unclear until recently. 
Nielsen and Chadha gave the inequality 
among the numbers of type-I and II NGMs 
and broken generators 
\cite{Nielsen:1975hm}. 
With the idea of Nambu \cite{Nambu:2004yia}, 
Watanabe and Brauner gave a conjecture 
in Ref.~\cite{PhysRevD.84.125013} 
stating that the number of type-II NGMs is 
a half the rank of
the Watanabe-Brauner (WB) matrix, whose components are commutators of 
generators corresponding to broken symmetries,  
sandwiched by the ground state. 
Then, the equality of the Nielsen-Chadha inequality 
and the Watanabe-Brauner conjecture 
have been proved recently by using 
 the effective Lagrangian approach based 
on a coset space \cite{PhysRevLett.108.251602}, 
by Mori's projection operator method \cite{Hidaka:2012ym},
and later by the Bogoliubov theory \cite{DTMN}.
Since this finding, extensive studies of 
NGMs in nonrelativistic systems 
have been made in various directions,    
such as massive (pseudo) NGMs 
\cite{Nicolis:2012vf,Nicolis:2013sga,
Watanabe:2013uya,Hayata:2014yga}, 
coupling to gauge fields 
\cite{Hama:2011rt,Gongyo:2014sra,Watanabe:2014qla}, 
space-time symmetry breaking 
\cite{Watanabe:2013iia,Hayata:2013vfa,Brauner:2014aha}, 
finite temperature and density \cite{Hayata:2014yga},  
higher derivative terms \cite{Andersen:2014ywa} 
and  
topological interaction \cite{Brauner:2014ata}. 
Furthermore, when there exists a topological soliton
or defect, NGMs are localized around it. 
Examples contain vortices in scalar BECs, 
helium superfluids \cite{Kobayashi:2013gba} 
and dense quark matter \cite{Eto:2013hoa}, 
a domain wall in ferromagnets \cite{Kobayashi:2014xua} 
and two-component BECs
\cite{PhysRevA.88.043612,Watanabe:2014zza}, 
and 
a skyrmion line in ferromagnets 
\cite{Watanabe:2014pea,Kobayashi:2014eqa}.
Among these cases, when zero modes are non-normalizable, 
there appear non-integer power dispersion 
relations, 
such as $\epsilon \propto k^{3/2}$ for 
a domain wall in two-component BECs
\cite{PhysRevA.88.043612} 
and $\epsilon \propto -k^2 \log k$ 
for a vortex in scalar BECs 
or helium superfluids 
\cite{Pitaevskii1961,Donnelly}. 
However, these dispersion relations become quadratic 
so they are type-II NGMs,
when transverse sizes are small enough 
as shown in Ref.~\cite{Kobayashi:2013gba,DTMN} 
for a vortex and in Ref.~\cite{DTMN} for a domain wall.
It has been also shown  in Ref.~\cite{DTMN}  that 
non-integer dispersion does not occur 
in the uniform ground states.

Among various approaches,
the effective Lagrangian 
based on coset spaces 
is very powerful because 
everything can be described in terms of 
only symmetry 
\cite{Coleman:1969sm,Callan:1969sn,
Leutwyler:1993gf,Watanabe:2014fva}.
However, it does not work in the presence of 
additional zero modes other than NGMs such as
quasi-NGMs \cite{Weinberg:1972fn,Georgi:1975tz}. 
This is the case that we discuss in this paper.

Quasi-NGMs appear when 
the symmetry of potential term or equation of motion 
is larger than the symmetry of Lagrangian or Hamiltonian 
and it is spontaneously broken 
in the ground state. 
In the mean field approximation, 
gapless modes are determined from 
the flat directions of the potential term, 
so that there can appear additional 
zero modes in addition to the conventional NGMs.
In relativistic theories, they appear
in technicolor models \cite{weinberg1996quantum} 
and supersymmetric field theories 
\cite{Kugo:1983ma,Lerche:1983qa, Shore:1984bh,
 Higashijima:1997ph,Nitta:1998qp,Higashijima:1999ki,
Nitta:2014fca}. 
When the Lagrangian 
in supersymmetric theories 
has a symmetry $G$, the 
superpotential 
always has an enlarged symmetry 
$G^{\mathbb C}$, a complexification of $G$. 
As a consequence, 
as proved in Refs.~\cite{Lerche:1983qa, Shore:1984bh},
there must appear at least 
one quasi-NGM when a global symmetry 
is spontaneously broken in supersymmetric theories
(in the absence of gauge interaction 
\cite{Higashijima:1999ki}). 
In nonrelativistic systems,
quasi-NGMs appear in condensed matter systems 
such as  
A-phase of $^3$He superfluids 
\cite{volovik2009universe} 
and $F=2$ spinor BECs 
\cite{PhysRevLett.105.230406},
and color superconductivity of dense quark matter 
\cite{Pang:2010wk}.

In our previous paper \cite{DTMN},
we presented 
the Bogoliubov theory approach 
to formulate general treatment of NGMs 
in nonrelativistic systems.
The advantages of this approach are 
that one can deal with 
additional zero modes such as quasi-NGMs 
in the same ground with NGMs
on one hand, 
and that one can also deal with NGMs for 
space-time symmetry breaking 
in the same manner on the other hand.

In this paper, we discuss quasi-NGMs 
in the Bogoliubov theory. 
In the presence of quasi-NGMs, 
there are two interesting physics 
that the effective field theory approach cannot 
deal with: 
\begin{enumerate}
\item
There can exist type-II quasi-NGMs 
consisting of one genuine zero mode 
and one quasi zero mode or two quasi zero modes.

\item
Some genuine zero modes can turn to 
quasi zero modes with keeping 
the total number of zero modes.
\end{enumerate}

Apparently, the effective Lagrangian based on 
coset space cannot deal with the first point
even if one ignores quasi-NGMs, because 
of type-II mode which contains only 
one symmetry generator. 
It is the same for the second point.

We focus on the cases that 
the potential term has non-compact symmetry
whose Lie algebra inevitably 
contains non-hermitian generators, 
which is motivated by quasi-NGMs in 
supersymmetric theories 
\footnote{In the case of the nematic phase 
of spin 2 BECs, the enhanced symmetry is described by a \textit{compact group} $ U(1)\times SO(5,\mathbb{R}) $, so the generators are all hermitian, and hence the problem explained here does not occur \cite{PhysRevLett.105.230406,DTMN}. 
}, 
and/or that the symmetry of the gradient term 
is reduced by 
multiple components with   
different particle masses.  
We show that the WB matrix does not work 
to count type-II modes
in this case.
On the other hand, we use the Gram matrix 
in the Bogoliubov theory. 
This reduces to the WB matrix only when 
all generators are hermitian. 
In general cases, we can still count the number of 
type-II modes by using the Gram matrix. 
We present the perturbation theory to 
calculate dispersion relations of (quasi-)NGMs. 
We find in general that there exist 
type-II modes made of two quasi-zero modes or one genuine and one quasi-zero modes, 
in addition to usual case of two genuine zero modes. 
We call the former quasi-NGMs and the latter 
conventional NGMs.
We demonstrate this theory by an explicit example 
exhibiting the above 
two features, that is, the complex linear $O(N)$ model 
\cite{Higashijima:2001yn}
consisting of $N$ complex scalar fields with 
$O(N)$ symmetry. 

We again point out that 
the coset space approach to the effective Lagrangian 
has a difficulty 
in this case.
Even when one includes quasi-NGMs in the 
effective theory, the coset space 
based on enlarged symmetry 
gives negative norm in general because of 
non-hermitian generators, 
resulting in the instability.
For instance, let us consider the simplest case that 
$U(1)^{\mathbb C}$ is 
spontaneously broken completely. 
Let $g = \exp i (\theta+ i R) %= r e^{i\theta}
\in U(1)^{\mathbb C}$ 
be a coset element 
where $\theta$ and $R$ are 
NG and quasi-NG modes, respectively. 
Then, the coset space ``Lagrangian" is
\begin{equation}
 {\cal L} = f^2 {\rm Re} (i g^{-1} \partial_{\mu}g)^2   
% = f^2 [(\partial_{\mu} \theta)^2 
% - {1\over r^2}(\partial_{\mu}r)^2 ]
 =  f^2 [(\partial_{\mu} \theta)^2 -  (\partial_{\mu} R)^2]
\end{equation}
where $R$, parameterizing a non-compact direction of 
$U(1)^{\mathbb C}$,  
has a negative norm.  
This is because we required 
an isometry of $U(1)^{\mathbb C}$ on the metric 
of the target space since  
in the coset approach one constructs a $G$-invariant 
metric on $G/H$.

Before closing introduction, 
we note that 
quasi-NGMs are different from 
pseudo-NGMs.
The latter appear when approximate symmetry 
of the Lagrangian is spontaneously broken, 
as the case of pions in the chiral symmetry breaking.
The effect of explicit symmetry breaking gives a mass gap 
to pseudo-NGMs even in the mean field approximation.
On the other hand, quasi-NGMs are gapless 
up to the mean field approximation.
However, quasi-NGMs may be gapped 
beyond the mean field approximation in general; 
in the perturbative regime where quantum effects are taken into account, they obtain a small gap, 
in which case quasi-NGMs become pseudo-NGMs.

This paper is organized as follows.
In Sec.~\ref{sec:general}, we give models and 
the Gross-Pitaevskii(-like) and Bogoliubov equations.
In Sec.~\ref{sec:formalism}, 
we give our general framework 
to obtain (quasi-)NGMs and their dispersion relations.
In Sec.~\ref{sec:example}, we give an example 
of the complex linear $O(N)$ model 
consisting of $N$ complex scalar fields with 
$O(N)$ symmetry, to demonstrate our theory.
Sec.~\ref{sec:summary} 
is devoted to a summary and discussion.
In Appendix \ref{app:perturbation}, 
we give detailed calculations 
for perturbation theory to obtain 
dispersion relations of (quasi-)NGMs.

%%%%%%%%%%%%%%%%%%%%%%%%%%
\section{The Model and Bogoliubov Equations}\label{sec:general}
	Here we construct a generalized theory of (quasi-)NGMs when the masses of kinetic terms are not necessarily equal to each other and/or the symmetry of the potential term is represented by a noncompact group. 
In such a situation, the counting by the WB matrix \cite{PhysRevD.84.125013} is no longer applicable
due to the non-hermitian properties of generators of a noncompact group, while the counting based on the Gram matrix \cite{DTMN} is still valid. 

\subsection{Model}\label{subsec:model}
	\indent For definiteness, we consider the following Hamiltonian describing 
the $N$-component scalar fields: 
	\begin{gather}
		\mathcal{H}=\mathcal{T}+\mathcal{V},\\
		\mathcal{T}=\frac{1}{2}\int\mathrm{d}x\left(2M_{ij}\nabla\psi_i^*\nabla\psi_j+L_{ij}\nabla\psi_i^*\nabla\psi_j^*+L_{ij}^*\nabla\psi_i\nabla\psi_j\right), \label{eq:nondiagonalT}\\
		\mathcal{V}=\int\mathrm{d}x F(\boldsymbol{\psi}^*,\boldsymbol{\psi}).
	\end{gather}
	Here, $ M_{ij}=M_{ji}^* $ and $ L_{ij}=L_{ji} $. The repeated indices imply a summation over those indices. 
Here and hereafter, we use the vectorial notation $ \boldsymbol{\psi}=(\psi_1,\dots,\psi_N)^T $, and $ F(\boldsymbol{\psi}^*,\boldsymbol{\psi}) $ is an abbreviation of  $ F(\psi_1^*,\dots,\psi_N^*,\psi_1,\dots,\psi_N) $.  The function $ F(\boldsymbol{\psi}^*,\boldsymbol{\psi}) $ is assumed to have the following symmetry
	\begin{align}
		F(\boldsymbol{\psi}^*,\boldsymbol{\psi})=F(g^*\boldsymbol{\psi}^*,g\boldsymbol{\psi}),
	\end{align}
	for $ {}^\forall g\in G_{\mathcal{V}} $, where the group $ G_{\mathcal{V}} $ is a subgroup of  $ GL(N,\mathbb{C}) $, which is not necessarily to be a compact group, and hence $ g $ need not be unitary. In order to guarantee the stability of the system, we require that the kinetic term $\mathcal{T}$ is always non-negative. 
This imposes the condition that the coefficient matrix
	\begin{align}
		\tilde{M}=\begin{pmatrix} M & L \\ L^* & M^* \end{pmatrix},\ M=M^\dagger, L=L^T,
	\end{align}
	must be positive-definite, where $ M $ and $ L $ are  $ N \times N $ matrices whose $ (i,j) $-components are given by $ M_{ij} $ and $ L_{ij} $. Since $ \tilde{M} $ is positive-definite, from the theorem of Ref.~\cite{Colpa1978}, there exist a symplectic transformation
	\begin{align}
		&\begin{pmatrix}\boldsymbol{\psi} \\ \boldsymbol{\psi}^* \end{pmatrix}=U\begin{pmatrix}\boldsymbol{\phi} \\ \boldsymbol{\phi}^* \end{pmatrix}, \\
		&U^{-1}=\sigma U^\dagger \sigma, \quad U=\tau U^* \tau, \label{eq:Bunitary} \\
		&\sigma=\begin{pmatrix} I_N & \\ & -I_N \end{pmatrix},\quad \tau=\begin{pmatrix} & I_N \\ I_N & \end{pmatrix}
	\end{align}
	such that  $ \tilde{M} $ is transformed into a diagonal matrix:
	\begin{align}
		U^\dagger \tilde{M}U=\operatorname{diag}\left( \frac{1}{2m_1},\dots,\frac{1}{2m_N},\frac{1}{2m_1},\dots,\frac{1}{2m_N} \right),\\
		\mathcal{T}=\int\mathrm{d}x\sum_{i=1}^N \frac{\nabla\phi_i^*\nabla\phi_i}{2m_i},\quad   
m_1,\dots,m_N>0 .
          \label{eq:diagonalT}
	\end{align}
Here, $ m_i $'s can be interpreted as particle masses 
of $N$-species.
	By positive-definiteness, the particle 
masses $ m_i $'s are all positive.

Here, in order to avoid confusions, we give a few remarks on terminologies and conventions. The matrix $ U $ satisfying Eq.~(\ref{eq:Bunitary}) is called \textit{``paraunitary''} in Refs.~\cite{Colpa1978,Colpa1986,Colpa1986II}, 
while it is called 
\textit{``Bogoliubov-unitary (B-unitary)''} in our work \cite{DTMN} since it represents a Bogoliubov transformation of bosonic field operators. The well-known symplectic transformation can be obtained by
	\begin{align}
		S=U_0^{-1}UU_0,\quad U_0=\frac{1}{\sqrt{2}}\begin{pmatrix} I_N & \mathrm{i}I_N \\ I_N&-\mathrm{i}I_N \end{pmatrix}.
	\end{align}
	Then, $ S $ is a real-valued matrix satisfying $ S^TJS=J $ with $ J=\sigma\tau $. See also Appendix B of Ref.~\cite{DTMN}.\\
	\indent In the diagonal form in Eq.~(\ref{eq:diagonalT}), if all masses  $ m_i $'s are different from each other,  $ \mathcal{T} $ is invariant only under the phase multiplication of each component  $ \phi_i\rightarrow \mathrm{e}^{\mathrm{i}\theta_i}\phi_i $, and hence the symmetry group of $ \mathcal{T} $, which henceforth we write as $ G_{\mathcal{T}} $, is given by $ G_{\mathcal{T}}=U(1)^N $. 
When some $m_i$ are degenerate, 
the symmetry group of $ \mathcal{T} $ is enhanced. 
For instance, 
if  $ m_1=m_2 $ but all remaining $ m_3,\dots, m_N $ are different,  $ G_{\mathcal{T}}=U(2)\times U(1)^{N-2} $. If all masses are the same, $ m_1=\dots=m_N $, the symmetry group is given by $ G_{\mathcal{T}}=U(N) $, 
which was treated in our previous work \cite{DTMN}. 
Most generally, if there are  $ p_i $ tuples consisting of $ N_i $ components with having the same mass, the symmetry group is given by
	\begin{align}
		G_{\mathcal{T}}=\prod_i U(N_i)^{p_i}, \quad  \sum_ip_iN_i=N.
	\end{align}
	Although we can always transform $ \mathcal{T} $ to the diagonal form in Eq.~(\ref{eq:diagonalT}), the choice of the field $ \phi_1,\dots,\phi_N $ which diagonalizes the kinetic term $ \mathcal{T} $ is not always convenient for consideration of the potential term $ \mathcal{V} $. Thus, henceforth, we construct a general theory with 
$ \mathcal{T} $ in the form of Eq.~(\ref{eq:nondiagonalT}). \\
	\indent For the potential term $ \mathcal{V} $, we allow it to have a symmetry of a noncompact group $ G_{\mathcal{V}} $. We emphasize that the total Hamiltonian $ \mathcal{H}=\mathcal{T}+\mathcal{V} $ only has a symmetry of a compact group $ G_{\mathcal{H}}=G_{\mathcal{T}}\cap G_{\mathcal{V}} $, since $ G_{\mathcal{T}} $ is a subgroup of the unitary group $ U(N) $.

The symmetry groups $ G_{\mathcal{T}} $ and $ G_{\mathcal{V}} $ of the kinetic term $ \mathcal{T} $ and the potential term $ \mathcal{V} $ generally have no inclusion relation, i.e., $ G_{\mathcal{T}} \not\subset G_{\mathcal{V}} $ and $ G_{\mathcal{V}} \not\subset G_{\mathcal{T}} $ may hold simultaneously. 
In this case, the Hamiltonian may have no continuous symmetry except for spacetime ones, {\it i.e.}
$ G_{\mathcal{H}}=\{e\} $, where $ \{e\} $ is a trivial group consisting only of an identity. 
It has no Noether conservation law except for energy and momentum. Even in this extreme case, 
there can exist gapless modes, i.e., quasi-NGMs, 
as we see below. This fact implies that the concepts of  Noether charges/currents are not indispensable in the formulation and proof of counting rule of NGMs and quasi-NGMs. Indeed, in our previous work \cite{DTMN}, the concept of symmetry was necessary only when we derive SSB-originated zero-modes and the conservation law was not used directly. 

%%%%%
\subsection{Gross-Pitaevskii and Bogoliubov equations}
	\indent Let us derive the fundamental equations and clarify the problem. The Hamilton equation describing the  $ N $-component order parameter $ \boldsymbol{\psi}=(\psi_1,\dots,\psi_N)^T $ is given by 
	\begin{align}
		\mathrm{i}\partial_t\psi_i&=-M_{ij}\nabla^2\psi_j-L_{ij}\nabla^2\psi_j^*+\frac{\partial F}{\partial \psi_i^*},\label{eq:GP01}\\
		-\mathrm{i}\partial_t\psi_i^*&=-M_{ij}^*\nabla^2\psi_j^*-L_{ij}^*\nabla^2\psi_j+\frac{\partial F}{\partial \psi_i}.\label{eq:GP02}
	\end{align}
	Borrowing the terms from condensed matter physics, we call the above equation as the Gross-Pitaevskii (GP) equation, though the current model does not necessarily describe the Bose-Einstein condensates. 
	Linearizing the GP equation, and writing the linearized fields as $ \delta\psi_i=u_i,\ \delta\psi_i^*=v_i $, we obtain
	\begin{align}
		\mathrm{i}\partial_tu_i&=-M_{ij}\nabla^2u_j-L_{ij}\nabla^2v_j+F_{ij}u_j+G_{ij}v_j, \label{eq:Bogo01} \\
		-\mathrm{i}\partial_tv_i&=-M_{ij}^*\nabla^2v_j-L_{ij}^*\nabla^2u_j+F_{ij}^*v_j+G_{ij}^*u_j \label{eq:Bogo02}
	\end{align}
	with
	\begin{align}
		F_{ij}=\frac{\partial^2 F}{\partial \psi_i^*\partial\psi_j},\quad G_{ij}=\frac{\partial^2 F}{\partial \psi_i^*\partial\psi_j^*}.
	\end{align}
	We also call Eqs.~(\ref{eq:Bogo01}) and (\ref{eq:Bogo02}) the Bogoliubov equation in accordance with condensed matter physics. Henceforth we write $ \boldsymbol{u}=(u_1,\dots,u_N)^T,\ \boldsymbol{v}=(v_1,\dots,v_N)^T $. Assuming the spacetime-independent $ \boldsymbol{\psi} $, and the plane-wave solution of the form $ (\boldsymbol{u},\boldsymbol{v})\propto \mathrm{e}^{\mathrm{i}(\boldsymbol{k}\cdot\boldsymbol{x}-\epsilon t)} $, we obtain the eigenvalue problem of the $ 2N\times 2N $ matrix:
	\begin{gather}
		\epsilon\begin{pmatrix}\boldsymbol{u} \\ \boldsymbol{v} \end{pmatrix}=(H_0+M_0k^2)\begin{pmatrix}\boldsymbol{u} \\ \boldsymbol{v} \end{pmatrix}, \label{eq:Bogostat}\\
		H_0=\begin{pmatrix}F & G \\ -G^* & -F^* \end{pmatrix},\quad M_0=\sigma\tilde{M}=\begin{pmatrix}M&L \\ -L^*&-M^*\end{pmatrix}, \label{eq:Bogostat2}
	\end{gather}
	where $ k=|\boldsymbol{k}| $, and $ F $ and $ G $ are the matrices whose  $ (i,j) $-components are given by $ F_{ij} $ and $ G_{ij} $, satisfying $ F=F^\dagger $ and $ G=G^T $. What we want to know is the dispersion relation $ \epsilon(k) $. We solve this problem by perturbation theory by regarding  $ H_0 $ as an unperturbed part and  $ M_0 $ as a perturbation term. If $ M_0=\sigma $, the problem reduces to the one which was solved in Ref.~\cite{DTMN}.

%%%%%%%%%%%%%%%%%%%%%%%%%%%%%%%%%%%%%%%%%
\section{General Theory of  (Quasi-)Nambu-Goldstone Modes} \label{sec:formalism}

\subsection{Conventional and quasi zero-mode solutions}\label{subsec:zeromode}
	The SSB-originated zero-mode solutions are the most important key concept in classification and perturbative calculations of dispersion relations of NGMs in the formulation by the Bogoliubov theory \cite{DTMN}. Here we generalize them for the case of quasi-NGMs. \\
	\indent First, let us consider the conventional SSB-originated zero-mode solutions derived from the symmetry of the total Hamiltonian $ G_{\mathcal{H}} $. Let $ \boldsymbol{\psi} $ be a solution of the GP equation (\ref{eq:GP01}) and (\ref{eq:GP02}), and let $ Q_j \ (j=1,\dots,n) $ be a generator of $ G_{\mathcal{H}} $ with $ n=\dim G_{\mathcal{H}} $. Since $ G_{\mathcal{H}} $ is a subgroup of the unitary group $ U(N) $, $ Q_j $ must be hermitian. We can immediately find the following property:
	\begin{align}
		& \boldsymbol{\psi} \text{ is a solution of the GP equation.} \nonumber \\
		\leftrightarrow \quad& \boldsymbol{\phi}=\mathrm{e}^{\mathrm{i}\alpha Q_j}\boldsymbol{\psi} \text{ is also a solution.}
	\end{align}
	Here $ \alpha $ is a real parameter. Then, differentiating the GP equation with substituted $ \boldsymbol{\phi} $ by $ \alpha $, and setting $ \alpha=0 $ after differentiation, we obtain the following particular solution for the Bogoliubov equation (\ref{eq:Bogo01}) and (\ref{eq:Bogo02}):
	\begin{align}
		\begin{pmatrix}\boldsymbol{u} \\ \boldsymbol{v} \end{pmatrix}=\boldsymbol{q}_j:=\begin{pmatrix} Q_j\boldsymbol{\psi} \\ -Q_j^*\boldsymbol{\psi}^* \end{pmatrix}, \quad j=1,\dots, n. \label{eq:SSBconventional ZM}
	\end{align}
	In particular, if we consider a time-independent $ \boldsymbol{\psi} $, we obtain the zero-energy solution of the Bogoliubov equation.  In order to distinguish them from that originated from the symmetry of $ G_{\mathcal{V}} $, henceforth we call them \textit{conventional zero-mode (conventional ZM) solutions}. (Here, in order to make the name short, we omit ``SSB-originated''.)  We note that if $ \boldsymbol{\psi} $ does not break the symmetry with respect to $ Q_j $, i.e.,  if $ \mathrm{e}^{\mathrm{i}\alpha Q_j}\boldsymbol{\psi}=\boldsymbol{\psi} $, Eq.~(\ref{eq:SSBconventional ZM}) only gives a zero vector. Therefore, if we write a number of broken symmetry as $ m(\le n) $, we obtain $ m $ linearly independent conventional ZMs. We also note that the conventional ZM solution exists even when $ \boldsymbol{\psi} $ has a spatial dependence, i.e., when it is written as $ \boldsymbol{\psi}=\boldsymbol{\psi}(\boldsymbol{r}) $. \\
	\indent Next, let us derive the zero-mode solutions originated from the symmetry of the potential term $ G_{\mathcal{V}} $. We henceforth call such solutions \textit{quasi-zero-mode (quasi-ZM) solutions}.  Let $ \boldsymbol{\psi}=(\psi_1,\dots,\psi_N)^T $ be a \textit{spacetime-independent} solution of the GP equation (\ref{eq:GP01}). Let $ \tilde{Q}_j \ (j=1,\dots,n') $ be a generator of $ G_{\mathcal{V}} $ but not that of $ G_{\mathcal{H}} $, where $ n'=\dim G_{\mathcal{V}}-\dim G_{\mathcal{H}} $. As already mentioned, $ \tilde{Q}_j $ need not be hermitian. Then, following the same argument with $ G_{\mathcal{H}} $, we can show
	\begin{align}
		& \boldsymbol{\psi} \text{ is a solution of the GP equation.} \nonumber \\
		\leftrightarrow \quad& \boldsymbol{\phi}=\mathrm{e}^{\mathrm{i}\alpha \tilde{Q}_j}\boldsymbol{\psi} \text{ is also a solution.}
\label{eq:prop-quasi-ZM}
	\end{align}
Also, by the same argument with conventional ZMs, we obtain the particular solution of the Bogoliubov equation
	\begin{align}
		\begin{pmatrix}\boldsymbol{u} \\ \boldsymbol{v} \end{pmatrix} = \tilde{\boldsymbol{q}}_j:=\begin{pmatrix} \tilde{Q}_j\boldsymbol{\psi} \\ -\tilde{Q}_j^*\boldsymbol{\psi}^* \end{pmatrix}, \quad j=1,\dots,n',
	\end{align}
	which we call a quasi-ZM. 

We note that the property in Eq.~(\ref{eq:prop-quasi-ZM})
holds \textit{only when $ \boldsymbol{\psi} $ does not have a spatial dependence}, 
because the kinetic term $ \mathcal{T} $ is not invariant under the symmetry operation of $ G_{\mathcal{V}} $. 
If the order parameter has a spatial dependence as $ \boldsymbol{\psi}(\boldsymbol{r}) $, then $ \boldsymbol{\phi}(\boldsymbol{r})=\mathrm{e}^{\mathrm{i}\alpha \tilde{Q}_j}\boldsymbol{\psi}(\boldsymbol{r}) $ is no longer a solution of the GP equation. This fact implies that the quasi-NGMs 
are expected to be fragile and are not robust against a perturbation inducing a spatial nonuniformity such as potential walls, vortices, and solitons.\\
	\indent At least in the systematic derivation of dispersion relations by perturbation theory, the distinction of the concept between conventional ZMs and quasi-ZMs is unimportant, as will be seen in the next subsection. \\
\subsection{Gram matrix and dispersion relations}\label{sec:grammatrix}
	\indent Let the linearly-independent conventional ZMs and quasi-ZMs derived in the previous subsection be $ \boldsymbol{q}_1,\dots,\boldsymbol{q}_m $ and $ \tilde{\boldsymbol{q}}_1,\dots,\tilde{\boldsymbol{q}}_{m'} $. For simplicity, we define $ \boldsymbol{q}_{m+l}=\tilde{\boldsymbol{q}}_l $ for $  l=1,\dots,m' $. Then, we introduce the Gram matrix $ P $ of size $ m+m' $, whose $ (i,j) $-component is given by 
	\begin{align}
		P_{ij}=(\boldsymbol{q}_i,\boldsymbol{q}_j)_\sigma, \label{eq:grammatrix}
	\end{align} 
	where the $ \sigma $-inner product is defined by \cite{DTMN}
	\begin{align}
		(\boldsymbol{x},\boldsymbol{y})_\sigma=\boldsymbol{x}^\dagger\sigma\boldsymbol{y},\quad \sigma=\begin{pmatrix} I_N & \\ & -I_N \end{pmatrix}
. \label{def:sigmaprod}
	\end{align}
	If $ (\boldsymbol{x},\boldsymbol{y})_\sigma=0 $,  $ \boldsymbol{x} $ and $ \boldsymbol{y} $ are said to be $ \sigma $-orthogonal.  If $ (\boldsymbol{x},\boldsymbol{x})_\sigma \ne 0 $,  $ \boldsymbol{x} $ is said to have finite norm. If not, it is said to have zero norm.\\ 
	\indent Let us block-diagonalize this Gram matrix. Since $ P $ is a pure-imaginary hermitian matrix, there exists a real orthogonal matrix $ O $ of size $ m+m' $ giving the following block-diagonal form:
	\begin{align}
		O^{-1}PO=(-\nu_1\sigma_y)\oplus\dotsb\oplus(-\nu_s\sigma_y)\oplus O_r,\quad \sigma_y=\begin{pmatrix}0&-\mathrm{i} \\ \mathrm{i} & 0\end{pmatrix}, \label{eq:gramblockdiag}
	\end{align}
	where $ r+2s=m+m' $ and $ \nu_1,\dots,\nu_s>0 $. Then the rank of $ P $ becomes
	\begin{align}
		\operatorname{rank}P=2s.
	\end{align}
	As shown below,  $ s $ gives the number of type-II gapless excitations. In the new basis giving this block-diagonal form in Eq.~(\ref{eq:gramblockdiag}), we write the first $ 2s $ vectors as $ \boldsymbol{x}_1^{(1)},\boldsymbol{x}_1^{(2)},\dots,\boldsymbol{x}_s^{(1)},\boldsymbol{x}_s^{(2)} $ and the rest $ r $ vectors as $ \boldsymbol{y}_1,\dots,\boldsymbol{y}_r $. Generally, they may be a linear combination of conventional ZMs and quasi-ZMs, i.e.,  $ \boldsymbol{q}_j $'s and $ \tilde{\boldsymbol{q}}_l $'s, and the mixing between conventional ZMs and quasi-ZMs can occur. 

 We can construct a finite-norm vector $ \boldsymbol{x}_i=\frac{1}{\sqrt{2\nu_i}}(\boldsymbol{x}_i^{(1)}-\mathrm{i}\boldsymbol{x}_i^{(2)}) $. These zero-mode solutions,  $ \boldsymbol{y}_1,\dots,\boldsymbol{y}_r $ and $ \boldsymbol{x}_1,\dots,\boldsymbol{x}_s $, become a seed of gapless excitations, i.e., a solution of the Bogoliubov equation Eq.~(\ref{eq:Bogostat}) with finite momentum $ k $ and the dispersion relation $ \epsilon(k) $  can be obtained by perturbation theory \cite{DTMN}. Since the calculation is a little long and complicated, we show this in Appendix~\ref{app:perturbation}. Here we only show the main result. \\

The zero-mode solutions introduced above satisfy
	\begin{align}
		(\boldsymbol{x}_i,\boldsymbol{x}_j)_\sigma&=\delta_{ij}, \label{eq:sigmabasis10} \\
		(\boldsymbol{y}_i,\boldsymbol{y}_j)_\sigma&=(\boldsymbol{y}_i,\boldsymbol{x}_j)_\sigma=0. \label{eq:sigmabasis11}
	\end{align} 
	While $ \boldsymbol{x}_i $'s have finite norm,  $ \boldsymbol{y}_i $'s have zero norm. All of them are  $ \sigma $-orthogonal to each other. Whether a given zero mode has finite or zero norm is crucial for classification of NGMs \cite{DTMN}. Let us assume that $ \sigma H_0 $ is positive-semidefinite and $ \sigma M_0 $ is positive-definite, where $ H_0 $ and $ M_0 $ are given in Eqs. (\ref{eq:Bogostat}) and (\ref{eq:Bogostat2}). This assumption ensures that the ground state has a linear stability \cite{DTMN}.  As we show in Appendix \ref{app:perturbation}, we can always find the following basis without changing the  $ \sigma $-orthogonal relations Eqs.~(\ref{eq:sigmabasis10}) and (\ref{eq:sigmabasis11}):
	\begin{align}
		(\boldsymbol{x}_i,M_0\boldsymbol{x}_j)_\sigma&=\frac{1}{\mu_i}\delta_{ij},\quad \mu_1,\dots,\mu_s>0, \\
		(\boldsymbol{y}_i,M_0\boldsymbol{y}_j)_\sigma&=2\kappa_i\delta_{ij},\quad \kappa_1,\dots,\kappa_r>0, \\
		(\boldsymbol{x}_i,M_0\boldsymbol{y}_j)_\sigma&=0.
	\end{align}
	Using this basis, we can perturbatively solve the Bogoliubov equation (\ref{eq:Bogostat}) with finite $ k\ne 0 $, and obtain the following result: The gapless mode arising from $ \boldsymbol{x}_i $ has a type-II dispersion relation
	\begin{align}
		\epsilon=\frac{1}{\mu_i}k^2+O(k^4),
	\end{align}
	and the gapless mode arising from $ \boldsymbol{y}_i $ has a type-I dispersion relation
	\begin{align}
		\epsilon=\sqrt{2\kappa_i}k+O(k^2).
	\end{align}
	Thus we have $ r $ type-I and $ s $ type-II gapless excitations, and the rank of $ P $ describes the number of type-II modes. See Appendix \ref{app:perturbation} for a more detailed and complete description.

	Now let us give a more precise definition for conventional and quasi- NGMs. As stated above, $ \boldsymbol{x}_i $'s and $ \boldsymbol{y}_i $'s are generally written as a linear combination of conventional ZMs  $ \boldsymbol{q}_1,\dots\boldsymbol{q}_m $ and quasi-ZMs $ \tilde{\boldsymbol{q}}_1,\dots,\tilde{\boldsymbol{q}}_{m'} $. If the zero mode solution $ \boldsymbol{y}_i $ is written by only using $ \boldsymbol{q}_j $'s, then a type-I gapless mode arising from  $ \boldsymbol{y}_i $ is called a type-I NGM. If  $ \boldsymbol{y}_i $ contains $ \tilde{\boldsymbol{q}}_j $'s, then the type-I gapless mode arising from  $ \boldsymbol{y}_i $ is called a type-I quasi-NGM. In the same way we define type-II NGMs and type-II quasi-NGMs depending on whether $ \boldsymbol{x}_i $ includes $ \tilde{\boldsymbol{q}}_j $'s or not. The classification explained here is summarized in Table \ref{tbl:clssfyqngs}. 
	\begin{table}[tb]
		\begin{center}
		\caption{\label{tbl:clssfyqngs}Classification of genuine and quasi- NGMs based on the properties of seed zero-mode solutions. $ \boldsymbol{q}_j $'s are conventional ZMs obtained from the symmetry of the Hamiltonian  $ G_{\mathcal{H}} $, and $ \tilde{\boldsymbol{q}}_j $'s are quasi-ZMs from the symmetry of the potential $ G_{\mathcal{V}} $ (See Subsec.\ref{subsec:zeromode}). A given gapless mode is a NGM  (quasi-NGM) if the seed zero-mode solution does not include (includes) quasi-ZMs in its linear combination. The dispersion relations are determined by the norm of zero-mode. The coefficients of type-II (quasi-)NGMs may be complex to make the norm finite.}
		{\small
	\begin{tabular}{r||c|c|}
	 & \begin{minipage}[c][3.5em][c]{10em} constituent of \\ seed zero mode \\ ($c_j,c_j'\in\mathbb{R},\ \alpha_j,\alpha_j'\in\mathbb{C}$.) \end{minipage}  & \begin{minipage}[c][3.5em][c]{7em} norm of \\ seed zero mode \\ $\vphantom{\alpha_j'}$ \end{minipage} \\
	\hline
	type-I NGM & $\boldsymbol{y}_i=\sum_jc_j\boldsymbol{q}_j$ & $(\boldsymbol{y}_i,\boldsymbol{y}_i)_\sigma=0$ \\
	\hline
	type-I quasi-NGM & $\boldsymbol{y}_i=\sum_jc_j\boldsymbol{q}_j+\sum_j c_j'\tilde{\boldsymbol{q}}_j$ & $(\boldsymbol{y}_i,\boldsymbol{y}_i)_\sigma=0$ \\
	\hline
	type-II NGM & $\boldsymbol{x}_i=\sum_j\alpha_j\boldsymbol{q}_j$ & $(\boldsymbol{x}_i,\boldsymbol{x}_i)_\sigma=1$ \\
	\hline
	type-II quasi-NGM & $\boldsymbol{x}_i=\sum_j\alpha_j\boldsymbol{q}_j+\sum_j \alpha_j'\tilde{\boldsymbol{q}}_j$ & $(\boldsymbol{x}_i,\boldsymbol{x}_i)_\sigma=1$ \\
	\hline
	\end{tabular}
		}
		\end{center}
	\end{table}

%%%%%%%%%%%%%%%%
\subsection{
The Gram matrix and the Watanabe-Brauner matrix}
Here we discuss the relation between 
the Gram matrix and the WB matrix \cite{PhysRevD.84.125013}, 
which are useful to count the number of type-II modes. \\
\indent When the generators of symmetry group are all hermitian, 
the Gram matrix is equivalent to the WB matrix:
	\begin{align}
		P_{ij}=\boldsymbol{\psi}^\dagger [Q_i,Q_j]\boldsymbol{\psi}\propto\rho_{ij}^{\text{WB}}.
	\end{align}
Therefore, both matrices work as well 
to count type-II modes. 
However, the generators of the noncompact group are not 
hermitian in general. 
	If some of generators are non-hermitian, we have 
	\begin{align}
		P_{ij}=\boldsymbol{\psi}^\dagger(Q_i^\dagger Q_j-Q_j^\dagger Q_i)\boldsymbol{\psi} \not\propto \rho_{ij}^{\text{WB}}.
	\end{align}
	Thus, it cannot be expressed as ``an expectation value of commutators''. 
In this case, the WB matrix is no longer equivalent to the Gram matrix and 
does not work anymore to count type-II modes.
Even in such the case, as demonstrated above, we can derive 
zero-mode solutions by differentiation with respect to parameters in the noncompact group, 
and can count the numbers of 
type-I and II modes by the Gram matrix 
in the same way with Ref.~\cite{DTMN}. \\
	\indent We note that if NGMs are classified based on not dispersion relations but whether conventional ZMs are paired (type-B) or unpaired (type-A) \cite{PhysRevLett.108.251602}, the criterion based on the WB matrix is still intact, though the dispersion relations cannot be predicted correctly.

%%%%%%%%%%%%%%%%%%%%%%%%%%%%%
\section{Example: complex linear $ O(N) $ Model}\label{sec:example}
In this section, 
we demonstrate the general theory given above 
by an explicit example, the complex linear $O(N)$ model. 
This model is also interesting in the point that 
it exhibits NGM-quasi-NGM changes, 
{\it i.e.}, some of NGMs change to quasi-NGMs 
in particular points in the target space, 
with preserving the total number of NGMs and quasi-NGMs.

\subsection{Complex linear $ O(N) $ model}
	Let us start with the complex $ O(N) $ model with the Lagrangian
	\begin{align}
		\mathcal{L}(\{\psi_i(x),\dot{\psi}_i(x)\})&=\int\mathrm{d}x\left(\frac{\mathrm{i}\psi_i^*\dot{\psi}_i-\mathrm{i}\dot{\psi}_i^*\psi_i}{2}\right)-\mathcal{T}-\mathcal{V},\\
		\mathcal{T}&=\int\mathrm{d}x\nabla\psi_i^*\nabla\psi_i,\\
		\mathcal{V}&=\int\mathrm{d}x F(\psi_i^*\psi_i^*,\psi_i\psi_i)
	\end{align}
	Here, the spatial dimension is arbitrary and the repeated indices imply the summation over $ 1\le i\le N $. The potential function $ F(s,s^*) $ is assumed to be real $ F(s,s^*)=F(s,s^*)^* $ and written only by the $O(N,\mathbb{C})$ singlet
	\begin{align}
		s:=\sum_{i=1}^N\psi_i\psi_i.
	\end{align}
	By this assumption, while the symmetry group of the total Lagrangian is $ G_{\mathcal{L}}=O(N,\mathbb{R}) $, the symmetry group of the potential term $ \mathcal{V} $ is $ G_{\mathcal{V}}=O(N,\mathbb{C}) $. The enhancement of the symmetry in the potential term is crucial for emergence of quasi-NGMs. The symmetry groups for each term and the total Lagrangian are summarized as
	\begin{align}
		G_{\mathcal{T}}&=U(N), \\
		G_{\mathcal{V}}&=O(N,\mathbb{C}), \label{eq:03GV} \\
		G_{\mathcal{L}}&=G_{\mathcal{T}}\cap G_{\mathcal{V}}=O(N,\mathbb{R}). \label{eq:03GH}
	\end{align}

Although we do not have to specify the form of 
the potential term, here we give two examples. 
The simplest example is given by
	\begin{align}
		F(s,s^*) = \lambda |s-r^2\mathrm{e}^{2\mathrm{i}\theta}|^2, \label{eq:O3pt01}
	\end{align}
where $ r $ and $ \lambda $ are positive and real, and $ \theta $ is real. 
A simple example of $F$ 
with an additional $U(1)$ symmetry, 
$ G_{\mathcal{V}}=U(1)\times O(N,\mathbb{C}) $,  is given by
	\begin{align}
		F(s,s^*)=|s|^4-2r^2|s|^2  \label{eq:O3pt02}
	\end{align}
with a real constant $r$. 

	In order to apply the general results obtained in the previous section, let us move on to the Hamiltonian formalism. The canonical momentum fields for $ \psi_i(x) $'s are given by
	\begin{align}
		\pi_i(x)=\frac{\delta \mathcal{L}}{\delta \dot{\psi}_i(x)}=\frac{\mathrm{i}\psi_i(x)^*}{2},\quad \pi_i(x)^*=\frac{\delta \mathcal{L}}{\delta \dot{\psi}_i(x)^*}=\frac{-\mathrm{i}\psi_i(x)}{2}.
	\end{align}
	Then, the Hamiltonian is introduced by the Legendre transformation, which coincides with $ \mathcal{T}+\mathcal{V} $: 
	\begin{align}
		\mathcal{H}&=\int\mathrm{d}x\left(\pi_i\dot{\psi}_i+\pi_i^*\dot{\psi}_i^*\right)-\mathcal{L}=\mathcal{T}+\mathcal{V}.
	\end{align}
	The symmetry of the Hamiltonian is the same with that of the Lagrangian: $ G_{\mathcal{H}}=G_{\mathcal{L}} $. The Hamilton equation for this system is 
	\begin{align}
		\mathrm{i}\partial_t\psi_i&=\frac{\delta \mathcal{H}}{\delta\psi_i^*}=-\nabla^2\psi_i+2\psi_i^*\left.\frac{\partial F(s^*,s)}{\partial s^*}\right|_{s=\psi_i\psi_i,\ s^*=\psi_i^*\psi_i^*}, \label{eq:O3GP01} \\
		-\mathrm{i}\partial_t\psi_i^*&=\frac{\delta \mathcal{H}}{\delta\psi_i}=-\nabla^2\psi_i^*+2\psi_i\left.\frac{\partial F(s^*,s)}{\partial s}\right|_{s=\psi_i\psi_i,\ s^*=\psi_i^*\psi_i^*}. \label{eq:O3GP02}
	\end{align}
	This is an analog of the GP equation describing Bose condensates, though the current system does not necessarily conserves a ``particle density''  $ \rho=\sum_i\psi_i^*\psi_i $ because of the absence of the $ U(1) $ symmetry. 
The potential term in Eq.~(\ref{eq:O3pt01}) is a case without  $ U(1) $-symmetry.
The particle density is conserved 
in the case with the $ U(1) $ symmetry, 
for instance for the potential term in 
Eq.~(\ref{eq:O3pt02}). \\
%%%%%%%%%%%%%%%%%%%%%%%%%%%%%%%%%%%%%%%%%
	\indent 
Next, we determine the ground state. 
Let us assume that the ground state of $ \psi_i $ is spatially uniform. Then, the ground state solely determined by the minimization of the potential $ \mathcal{V} $.  From Eqs.~(\ref{eq:O3GP01}) and (\ref{eq:O3GP02}),  $ \frac{\partial F}{\partial s}=\frac{\partial F}{\partial s^*}=0 $ hold in the stationary state. 

	We can generally show that any  $N$-component complex vector $ \boldsymbol{\psi}=(\psi_1,\cdots,\psi_N)^T $ can be transformed into the following form by $ O(N,\mathbb{R}) $ transformation:
	\begin{align}
		\boldsymbol{\psi}=r\mathrm{e}^{\mathrm{i}\theta}\begin{pmatrix} \cosh\varphi \\ \mathrm{i}\sinh\varphi \\ 0 \\ \vdots \\ 0 \end{pmatrix}, \label{eq:stdfrmphi}
	\end{align}
	where $ r,\theta,\varphi \in\mathbb{R} $ and $ r>0,\ \varphi>0 $. Thus, without loss of generality, we assume that the solution of Eqs.~(\ref{eq:O3GP01}) and (\ref{eq:O3GP02}) is given with Eq.~(\ref{eq:stdfrmphi}). 
Note that the singlet $ s $ is given by
	\begin{align}
		s=\psi_i\psi_i=r^2\mathrm{e}^{2\mathrm{i}\theta},
	\end{align}
	which does not depend on $ \varphi $. Therefore, the order parameter space consisting of ground states has a residual degree of freedom represented by $ \varphi $, in addition to the NGM degree of freedom due to  $ O(N,\mathbb{R}) $-rotation symmetry. This degree of freedom is directly related to the emergence of quasi-NGMs. We can further understand it by an enhanced group symmetry $ G_{\mathcal{V}} $ as follows.%discussed in the next subsection.\\

When we use $G_{\mathcal{V}} = O(N,\mathbb{C}) $, 
$ \boldsymbol{\psi} $ can be transformed to 
\begin{align}
		\boldsymbol{\psi}=r\mathrm{e}^{\mathrm{i}\theta}
\begin{pmatrix} 1\\ 0 \\ \vdots \\ 0 \end{pmatrix}, \label{eq:stdfrmphi0}
	\end{align}
that is, $\varphi$ can be taken to be zero. 
The unbroken symmetry is then 
$H_{\mathcal{V}} = O(N-1,\mathbb{C}) $, and 
the order parameter manifold is 
\begin{align}
 {G_{\mathcal{V}} \over H_{\mathcal{V}} }= { O(N,\mathbb{C}) \over O(N-1,\mathbb{C})} 
\simeq
 T^* \left[{ O(N,\mathbb{R}) \over O(N-1,\mathbb{R})} 
\right]
\simeq T^* S^{N-1} .\label{eq:OPS}
\end{align}
Since the gradient term is invariant only 
under $O(N,\mathbb{R})$,
this space does not have an $O(N,\mathbb{C})$ isometry 
but only an $O(N,\mathbb{R})$ isometry.
The unbroken 
symmetry $H_{\mathcal{L}}$ of Lagrangian 
is not unique, depending on $\varphi$. It is 
	\begin{align}
		H_{\mathcal{L}}=\begin{cases}O(N-1,\mathbb{R}) & \text{for }\ \varphi=0,  \\ O(N-2,\mathbb{R}) & \text{for }\ \varphi\ne0. \end{cases}
	\end{align}
Therefore, the number of NGMs varies 
depending on $\varphi$.
This can be understood by noting that 
the unbroken symmetry 
$H_{\mathcal{V}}{}_{\varphi}$ depends on $\varphi$ 
as  
$H_{\mathcal{V}}{}_{\varphi} 
= g H_{\mathcal{V}}{}_{\varphi=0} g^{-1}$ 
with $g \in G_{\mathcal{V}}$ 
and the unbroken symmetry of the potential, 
 $H_{\mathcal{V}}{}_{\varphi}$,  
at each $\varphi$ is isomorphic to each other, 
while the unbroken symmetry of Lagrangian, 
\begin{align}
H_{\mathcal{L}} = H_{\mathcal{V}}{}_{\varphi} \cap U(N),
\end{align}
does not have to be isomorphic to each other 
for every $\varphi$.

When the manifold in Eq.~(\ref{eq:OPS})
is endowed with a Ricci-flat K\"ahler metric,
it is the Eguchi-Hanson space \cite{Eguchi:1980jx} 
for $N=3$, 
the deformed conifold 
\cite{Candelas:1989js} 
for $N=4$, 
and the Stenzel metric 
\cite{Stenzel:1993,Higashijima:2001yn}
for general $N$.

%%%%%%%%%%%%%%%%%%%%%%%%%%%%%%%%%%%%%%
\subsection{The Bogoliubov equation}
%%%%%%%%%%%%%%%%%%%%%%%%%%%%%%
\indent  The linearization of the GP equation yields the Bogoliubov equation. That is, substituting  $ (\psi_i,\psi_i^*)=(\psi_i+\delta\psi_i,\psi_i^*+\delta\psi_i^*) $ to Eqs. (\ref{eq:O3GP01}) and (\ref{eq:O3GP02}) and ignoring the higher-order terms w.r.t.  $ \delta\psi_i $'s and $ \delta\psi_i^* $'s and rewriting $ (\delta\psi_i,\delta\psi_i^*)=(u_i,v_i) $,  we get
	\begin{align}
		\mathrm{i}\partial_t u_i&=-\nabla^2u_i+4\frac{\partial^2F}{\partial s\partial s^*}\psi_i^*\psi_ju_j+\left( 2\frac{\partial F}{\partial s^*}\delta_{ij}+4\frac{\partial^2 F}{\partial s^{*2}}\psi_i^*\psi_j^* \right)v_j,\\
		-\mathrm{i}\partial_tv_i&=-\nabla^2v_i+4\frac{\partial^2F}{\partial s\partial s^*}\psi_i\psi_j^*v_j+\left( 2\frac{\partial F}{\partial s}\delta_{ij}+4\frac{\partial^2F}{\partial s^2}\psi_i\psi_j \right)u_j,
	\end{align}
	where the notations of substitution $ |_{s=\psi_i\psi_i,\ s^*=\psi_i^*\psi_i^*} $  for derivatives of $ F $ are omitted.\\
%%%%%%%%%%%%%%%%%%%
	\indent Then the stationary Bogoliubov equation with an eigenenergy $ \epsilon $ can be obtained by substitution $ (u_i,v_i)\propto \mathrm{e}^{\mathrm{i}(\boldsymbol{k}\boldsymbol{x}-\epsilon t)} $, yielding
	\begin{align}
		\epsilon\begin{pmatrix}\boldsymbol{u} \\ \boldsymbol{v}\end{pmatrix}=\begin{pmatrix}F+k^2 & G \\ -G^* & -F^*-k^2 \end{pmatrix}\begin{pmatrix}\boldsymbol{u} \\ \boldsymbol{v}\end{pmatrix}, \label{eq:03Bogostt}
	\end{align}
	where $ \boldsymbol{u}=(u_1,\dots,u_N)^T $ and $ \boldsymbol{v}=(v_1,\dots,v_N)^T $ and $ F $ and $ G $ are $ N\times N $ matrices whose components are given by
	\begin{align}
		F_{ij}=4\frac{\partial F}{\partial s\partial s^*}\psi_i^*\psi_j,\quad G_{ij}=4\frac{\partial^2F }{\partial s^{*2}}\psi_i^*\psi_j^*.
	\end{align}

%%%%%%%%%%%%%%
	\indent Henceforth, for simplicity, we concentrate on the case of $ O(3) $ model. However, the essence is the same for general $ N $. 
When $ \psi_i $ is given by Eq.(\ref{eq:stdfrmphi}), the matrices in Eq.~(\ref{eq:03Bogostt}) reduce to
	\begin{align}
		F=4r^2\frac{\partial^2F }{\partial s\partial s^*}\begin{pmatrix} \cosh^2\varphi & \mathrm{i}\cosh\varphi\sinh\varphi & 0 \\ -\mathrm{i}\cosh\varphi\sinh\varphi & \sinh^2\varphi & 0 \\ 0&0&0 \end{pmatrix},\\
		G=4r^2\mathrm{e}^{-2\mathrm{i}\theta}\frac{\partial^2F}{\partial s^{*2}}\begin{pmatrix} \cosh^2\varphi & -\mathrm{i}\cosh\varphi\sinh\varphi & 0 \\ -\mathrm{i}\cosh\varphi\sinh\varphi & -\sinh^2\varphi & 0 \\ 0&0&0 \end{pmatrix}.
	\end{align}
	Solving the Bogoliubov equation (\ref{eq:03Bogostt}), we soon find the following dispersion relations:
	\begin{align}
		\epsilon&=k^2 \quad \text{(doubly degenerate)}, \label{eq:03typeIIdb} \\
		\epsilon&=\left[16(F_{ss^*}^2-F_{ss}F_{s^*s^*})r^4\cosh^2(2\varphi)\right. \nonumber \\
		&\qquad\qquad\qquad\left.+8F_{ss^*}r^2\cosh(2\varphi)k^2+k^4\right]^{1/2}. \label{eq:03gapful}
%		&\epsilon=\sqrt{16(F_{ss^*}^2-F_{ss}F_{s^*s^*})r^4\cosh^2(2\varphi)+8F_{ss^*}r^2\cosh(2\varphi)k^2+k^4} \label{eq:03gapful}
	\end{align}
	Here, $ F_{ss^*}=\frac{\partial^2F }{\partial s\partial s^*},\ F_{ss}=\frac{\partial^2F }{\partial s^2}, $ and $ F_{s^*s^*}=\frac{\partial^2F }{\partial s^{*2}} $ and we have only shown the positive dispersion relations. Thus, we have two type-II and one gapful excitations. 

The gapful mode given in Eq.~(\ref{eq:03gapful}) becomes 
a type-I mode, when the relation 
	\begin{align}
		F_{ss^*}^2-F_{ss}F_{s^*s^*}=0 \label{eq:03U1cond00}
	\end{align}
holds. This corresponds to the emergence of the $ U(1) $-symmetry as follows; If  $ F(s,s^*) $ is a function depending only on $ |s|^2 $, i.e., if $ F $ can be written as $ F(s,s^*)=\tilde{F}(|s|^2) $, the potential is also invariant under the $U(1)$ transformation $ \boldsymbol{\psi}\rightarrow \mathrm{e}^{\mathrm{i}\theta}\boldsymbol{\psi} $ and $ G_{\mathcal{V}} $ becomes $ G_{\mathcal{V}}=U(1)\times O(3,\mathbb{C}) $. In this case, the following holds:
	\begin{align}
		s\frac{\partial F}{\partial s}=s^*\frac{\partial F}{\partial s^*}=|s|^2 \tilde{F}(|s|^2). \label{eq:03U1cond}
	\end{align}
	Differentiating Eq.~(\ref{eq:03U1cond}) by $ s $ and $ s^* $ and using the stationary condition $ \frac{\partial F}{\partial s}=\frac{\partial F}{\partial s^*}=0 $, we have
	\begin{align}
		\frac{\partial^2F }{\partial s^2}=\frac{s^*}{s}\frac{\partial^2F }{\partial s\partial s^*},\quad \frac{\partial^2F }{\partial s^{*2}}=\frac{s}{s^*}\frac{\partial^2F }{\partial s\partial s^*},
	\end{align}
	which leads Eq. (\ref{eq:03U1cond00}). Thus, the emergence of the type-I mode can be explained by the emergence of the $ U(1) $ symmetry. \\
	\indent The above result for general potential $ F(s,s^*) $ can be checked by the 
specific examples of the potential terms given in Eqs.~(\ref{eq:O3pt01}) and (\ref{eq:O3pt02}). 
	In the next subsection, we investigate conventional ZMs and quasi-ZMs and identify the origin of the type-II modes, 
given in Eq.~(\ref{eq:03typeIIdb}). 

%%%%%%%%%%%%
\subsection{Zero-mode solutions}
	Let us apply the result of Subsec.~\ref{subsec:zeromode} to the current model. The symmetry of the total Lagrangian or Hamiltonian is given by Eq.~(\ref{eq:03GH}). $ G_{\mathcal{L}}=G_{\mathcal{H}}=O(3,\mathbb{R}) $ has generators $ T_1,\, T_2, $ and $ T_3 $, where  $ T_i $ is a generator of rotation with respect to $ i $-axis, and its components are given by  $ (T_i)_{jk}=-\mathrm{i}\epsilon_{ijk} $ with $ \epsilon_{ijk} $ being the Levi-Civita tensor. The symmetry of the potential is given by Eq.~(\ref{eq:03GV}).  $ G_{\mathcal{V}}=O(3,\mathbb{C}) $ is six-dimensional and the generators are given by $ \mathrm{i}T_1,\,\mathrm{i}T_2, $ and $ \mathrm{i}T_3 $ in addition to those of $ G_{\mathcal{L}} $. 
Thus, we have at most  six zero-mode solutions:
	\begin{align}
		\begin{pmatrix}\boldsymbol{u}\\ \boldsymbol{v} \end{pmatrix}=\begin{pmatrix}Q\boldsymbol{\psi} \\ -Q^*\boldsymbol{\psi}^*\end{pmatrix},\quad Q=T_1,T_2,T_3,\mathrm{i}T_1,\mathrm{i}T_2, \text{ and } \mathrm{i}T_3. \label{eq:03zeromodes}
	\end{align}
	These are the solutions of the Bogoliubov equation Eq.~(\ref{eq:03Bogostt}) with $\epsilon = 0$ for $ k=0 $. If $ Q $ is a linear combination of $ T_1,T_2,T_3 $, then the zero mode solution becomes a conventional ZM. If $ \mathrm{i}T_1,\mathrm{i}T_2,\mathrm{i}T_3 $ are included, it becomes a quasi-ZM. Any state $ \boldsymbol{\psi} $ represented by Eq.~(\ref{eq:stdfrmphi}) 
preserves $ H_{\mathcal{V}}=O(2,\mathbb{C}) $ unbroken symmetry, because
	\begin{align}
		\alpha(\cosh\varphi T_1+\mathrm{i}\sinh\varphi T_2)\boldsymbol{\psi}=0,\quad \alpha \in \mathbb{C}. \label{eq:symofpsi}
	\end{align}
	So, the number of broken continuous symmetry in $ G_{\mathcal{V}} $ is four and there are only four linearly-independent solutions in Eq. (\ref{eq:03zeromodes}). Whether Eq.~(\ref{eq:symofpsi}) includes the symmetry within $ G_{\mathcal{L}} $ or not depends on the value of $ \varphi $. If $ \varphi\ne0 $, two elements in Eq.~(\ref{eq:symofpsi}) are non-hermitian and it has no symmetry operation in  $ G_{\mathcal{L}} $, and hence  $ H_{\mathcal{L}}=\{e\} $. On the other hand, if $ \varphi=0 $, it has a hermitian element $ T_1 $ and $ H_{\mathcal{L}}=O(2,\mathbb{R}) $. Thus, the numbers of conventional ZMs and quasi-ZMs change depending on whether $ \varphi=0 $ or not, with keeping the total number of zero modes. \\
	\indent If $ \varphi\ne0 $, we have three conventional ZMs
	\begin{align}
		\boldsymbol{q}_i=\begin{pmatrix} T_i\boldsymbol{\psi} \\ -T_i^*\boldsymbol{\psi}^* \end{pmatrix},\quad i=1,2,3,
	\end{align}
	and one quasi-ZM
	\begin{align}
		\tilde{\boldsymbol{q}}_3=\begin{pmatrix} \mathrm{i}T_3\boldsymbol{\psi} \\ \mathrm{i}T_3^*\boldsymbol{\psi}^* \end{pmatrix}.
	\end{align}
	The other modes written by $ \mathrm{i}T_1 $ and $ \mathrm{i}T_2 $ are not independent of those of $ T_1 $ and $ T_2 $. We remark that the quasi-ZM $ \tilde{\boldsymbol{q}}_3 $ can be also obtained by differentiation by a parameter $ \varphi $, i.e., $\tilde{\boldsymbol{q}}_3 \propto \partial_\varphi(\boldsymbol{\psi},\boldsymbol{\psi}^*)^T $.  From them, we can construct finite-norm vectors as
	\begin{align}
		\boldsymbol{x}_1&=\frac{1}{2r\sinh\varphi}\boldsymbol{q}_1-\frac{\mathrm{i}}{2r\cosh\varphi}\boldsymbol{q}_2 \nonumber \\
		&=(0,0,\mathrm{e}^{\mathrm{i}\theta},0,0,0)^T,\\
		\boldsymbol{x}_2&=\frac{\boldsymbol{q}_3-\mathrm{i}\tilde{\boldsymbol{q}}_3}{2r} \nonumber \\
		&=(\sinh\varphi\mathrm{e}^{\mathrm{i}\theta},\mathrm{i}\cosh\varphi\mathrm{e}^{\mathrm{i}\theta},0,0,0,0)^T.
	\end{align}
	These zero-mode solutions give rise to to type-II modes, 
if we solve the equation Eq.~(\ref{eq:03Bogostt}) with $ k\ne 0 $ perturbatively, as shown in Subsec.~\ref{sec:grammatrix} and Appendix \ref{app:perturbation}. Since $ \boldsymbol{x}_1 $ can be written by a linear combination of conventional ZMs, the type-II mode arising from $ \boldsymbol{x}_1 $ is a conventional NGM. On the other hand,  $ \boldsymbol{x}_2 $ is a linear combination of a conventional ZM and quasi-ZM, and hence the type-II mode arising from $ \boldsymbol{x}_2 $ is a quasi-NGM. We thus obtain the two type-II modes in Eq.~(\ref{eq:03typeIIdb}) from zero-mode analysis, and identified one to be a genuine type-II 
NGM and the other to be a quasi-NGM made of 
one conventional ZM and one quasi-ZM. \\
	\indent Next, let us consider the case $ \varphi=0 $. In this case, since $ T_1\boldsymbol{\psi}=\boldsymbol{0} $, the number of conventional ZMs is two:
	\begin{align}
		\boldsymbol{q}_i=\begin{pmatrix} T_i\boldsymbol{\psi} \\ -T_i^*\boldsymbol{\psi}^* \end{pmatrix},\quad i=2,3.
	\end{align}
	Instead, we have two quasi-ZMs:
	\begin{align}
		\tilde{\boldsymbol{q}}_i=\begin{pmatrix} \mathrm{i}T_i\boldsymbol{\psi} \\ \mathrm{i}T_i^*\boldsymbol{\psi}^* \end{pmatrix},\quad i=2,3.
	\end{align}
	The finite-norm eigenvectors are given by
	\begin{align}
		\boldsymbol{x}_1=\frac{\boldsymbol{q}_2-\mathrm{i}\tilde{\boldsymbol{q}}_2}{2r} %\nonumber \\
		&=(0,0,-\mathrm{i}\mathrm{e}^{\mathrm{i}\theta},0,0,0)^T,\\
		\boldsymbol{x}_2=\frac{\boldsymbol{q}_3-\mathrm{i}\tilde{\boldsymbol{q}}_3}{2r} %\nonumber \\
		&=(0,\mathrm{i}\mathrm{e}^{\mathrm{i}\theta},0,0,0,0)^T.
	\end{align}
	Both the modes are written as a linear combination of a conventional ZM and quasi-ZM, thus the two type-II modes in Eq.~(\ref{eq:03typeIIdb}) are both quasi-NGMs.

While we have concentrated on the complex $O(3)$ model,
the analysis can be easily 
extended to the complex $O(N)$ model.
At $\varphi =0$, there are $N-1$ type-II quasi-NGMs 
consisting of $N-1$ conventional ZMs 
and $N-1$ quasi-ZMs, 
and at $\varphi \neq 0$, 
there are $2N-3$ conventional ZMs
and one quasi-ZM, yielding $ N-2 $ type-II NGM and one type-II quasi-NGM.
With the $U(1)$ symmetric potential such as Eq.~(\ref{eq:O3pt02}), 
there is also one type-I NGM. These are summarized in Table \ref{ta:qngON}.

\begin{table*}[tb]
	\begin{center}
	\caption{\label{ta:qngON} The numbers of conventional ZMs, quasi-ZMs, type-II NGMs and quasi-NGMs in the complex linear $ O(N) $ model for the cases $ \varphi=0 $ and $ \varphi\ne0 $ in Eq.~(\ref{eq:stdfrmphi}). Here we assume that $ G_{\mathcal{V}} $ does not have a $ U(1) $-symmetry.}
	{\small
	\begin{tabular}{r||c|c|c|c|c|c|}
	 &  $ H_{\mathcal{L}} $ & $H_{\mathcal{V}}$ & \begin{minipage}[c][2em][c]{8em} \#  of \\  conventional ZMs \end{minipage} & \begin{minipage}[c][2em][c]{5em} \# of \\  quasi-ZMs \end{minipage} & \begin{minipage}[c][2em][c]{6em} \# of \\ type-II NGMs \end{minipage} & \begin{minipage}[c][2em][c]{10em} \# of \\ type-II quasi-NGMs \end{minipage} \\
	\hline
	 $ \varphi=0 $  & $O(N-1,\mathbb{R})$ & $O(N-1,\mathbb{C})$ & $N-1$ & $N-1$&$0$&$N-1$ \\
	\hline
	 $ \varphi\ne0 $  & $O(N-2,\mathbb{R})$ & $O(N-1,\mathbb{C})$ & $2N-3$ &$1$&$N-2$&$1$ \\
	\hline
	\end{tabular}
	}
	\end{center}
\end{table*}
%%%%%%%%%%%%%%%
\section{Summary and discussion} \label{sec:summary}
We have presented a framework 
in the Bogoliubov theory 
to study 
NGMs and quasi-NGMs 
 in the same ground.
We have found two  phenomena 
of quasi-NGMs that 
the effective Lagrangian approach based on 
coset spaces cannot deal with. 
There exist two kinds of type-II 
gapless 
modes with quadratic dispersion relations,
a genuine NGM consisting of two conventional ZMs
and a quasi-NGM consisting of one conventional ZM 
and one quasi-ZM or two quasi-ZMs. 
Depending on the moduli, 
genuine NGMs can change 
into quasi-NGMs with preserving 
the total number of 
gapless modes. 
We have discussed the cases that 
the potential term has non-compact symmetry, 
whose Lie algebra inevitably 
contains non-hermitian generators,  
and/or that the symmetry of the gradient term 
is reduced. 
We have shown that the WB matrix 
can count only NGMs, 
while the Gram matrix in our framework 
can count both NGMs and quasi-NGMs. 
We have presented perturbation theory to 
obtain dispersion relations. 
We have demonstrated the theory by 
the complex linear $O(N)$ model 
consisting of $N$ complex scalar fields with 
$O(N)$ symmetry.

Some comments on quasi-NGMs 
are addressed here. 
Quasi-NGMs can be also localized 
in the vicinity a topological soliton.
An example can be found  in 
a baby Skyrmion line \cite{Kobayashi:2014eqa}.
In this case, 
dilatation
and $U(1)$ phase rotation 
are symmetries of equations of motion 
and of Lagrangian, respectively.
They are 
spontaneously broken in the presence 
of the baby Skyrmion,  
and
a type-II NGM, dilaton-magnon, 
consisting of quasi ZM (the dilatation)
and conventional ZM  (the $U(1)$ phase) 
is localized around it.

We have obtained quasi-NGMs 
within the framework of the mean field approximation.
However, beyond mean field approximation 
quasi-NGMs are fragile  against 
quantum corrections 
and will be gapped 
because the gradient (kinetic) term 
is not invariant under the 
enlarged symmetry of the potential, 
while genuine type-II NGMs remain gapless 
in quantum corrections 
even in lower dimensions \cite{Nitta:2013wca}.
It will be important 
to study the fate of type-II modes 
consisting of one conventional ZM 
and one quasi-ZM 
under quantum corrections.
When the quasi-ZM is gapped by 
quantum corrections, such a type-II mode 
may change to a type-I NGM.
This was demonstrated in the context 
of a Skyrmion line \cite{Kobayashi:2014eqa}, 
where a coupled dilation-magnon appears as 
a type-II quasi-NG mode. 
If we add an explicit breaking term for 
the dilatational symmetry 
(which mimics quantum corrections  
beyond the mean field approximation), 
the dilaton is gapped and 
the magnon becomes a type-I NG mode.

Quasi-NGMs are also fragile against  
spatial (or temporal) gradients
because of the same reason. 
Quasi-NGMs in the bulk may be gapped 
for instance in the vicinity of a topological soliton.
Detailed discussion on this direction 
remains as a future problem.

%%%%%%%%%%%%
\section*{Acknowledgments}
The work of MN is supported in part by Grant-in-Aid for Scientific
Research (No. 25400268) and by the ``Topological Quantum Phenomena'' 
Grant-in-Aid for Scientific Research on Innovative Areas (No. 25103720)  
from the Ministry of Education, Culture, Sports, Science and Technology 
(MEXT) of Japan. 

\appendix

\makeatletter
\renewcommand{\theequation}{\Alph{section}.\arabic{equation}}
\@addtoreset{equation}{section}
\makeatother

\section{Perturbation theory}\label{app:perturbation}
	In this appendix, we present a perturbation theory for the matrix of the Bogoliubov equation $ H_0+M_0k^2 $ [Eq.~(\ref{eq:Bogostat})]. We solve the eigenvalue problem of this matrix by regarding $ H_0 $ as an unperturbed part and $ M_0k^2 $ as a perturbation term, with knowing the zero-energy eigenvectors of $ H_0 $, i.e., conventional ZMs and quasi-ZMs derived in Subsec.~\ref{subsec:zeromode}. \\
	\indent If $ M_0=\sigma $, this problem reduces to our previous work \cite{DTMN}. Thus, the content in this appendix gives a generalization of a perturbation theory when the perturbation term $ M_0 $ is a more general Bogoliubov-hermitian matrix.\\
	\indent Here we introduce a few terminologies from Ref.~\cite{DTMN}. The Bogoliubov-unitary matrix is already defined in the main text [Subsec.~\ref{subsec:model}, Eq~.(\ref{eq:Bunitary})]. If a matrix $ H $ satisfy the following condition,  $ H $ is called Bogoliubov-hermitian (B-hermitian):
	\begin{align}
		H^\dagger=\sigma H \sigma,\quad H=-\tau H^*\tau.
	\end{align}
	Both $ H_0 $ and $ M_0 $ in Eq.~(\ref{eq:Bogostat}) are B-hermitian. Several linear-algebraic properties for B-hermitian and B-unitary matrices are summarized in Sec. 3 of Ref.~\cite{DTMN}. Here we extract only a few practically-important properties:
	\begin{itemize}
		\item If $ \boldsymbol{w} $ is a right eigenvector of $ H $ with a real eigenvalue $ \lambda $,  $ \tau \boldsymbol{w}^* $ is a right eigenvector of $ H $ with eigenvalue $ -\lambda $. Thus, positive and negative eigenvalues always appear in pairs.
		\item An analog of self-adjointness: $(\boldsymbol{x},H\boldsymbol{y})_\sigma=(H\boldsymbol{x},\boldsymbol{y})_\sigma$.
		\item If we write a B-unitary matrix $ U $ as an array of column vectors $ U=(\boldsymbol{x}_1,\dots,\boldsymbol{x}_N,\tau\boldsymbol{x}_1^*,\dots,\tau\boldsymbol{x}_N^*) $, these $ 2N $ vectors are linearly-independent and $ \sigma $-orthogonal to each other.
	\end{itemize}
	\indent First we derive a Colpa's standard form \cite{Colpa1986} for $H_0$. Let us assume that $ H_0 $ is a B-hermitian matrix such that $ \sigma H_0 $ is positive-semidefinite, and the eigenvectors of $ H_0 $ with zero eigenvalue are exhausted by $ \boldsymbol{y}_1,\dots,\boldsymbol{y}_r,\boldsymbol{x}_1,\dots,\boldsymbol{x}_s,\tau\boldsymbol{x}_1^*,\dots,\boldsymbol{x}_s^* $, which are derived in Subsec.~\ref{sec:grammatrix}. Following the result by Colpa \cite{Colpa1986} (See also Sec. 3 of Ref.~\cite{DTMN}), for each $\boldsymbol{y}_i$, there exists a unique generalized eigenvector $ \boldsymbol{z}_i $ satisfying the relations $ H_0\boldsymbol{z}_i=2\boldsymbol{y}_i,\ (\boldsymbol{y}_i,\boldsymbol{z}_j)_\sigma=2\delta_{ij} $ \cite{Colpa1986}. We also write the eigenvector with the positive eigenvalue $ \lambda_i $ as $ \boldsymbol{w}_i $,\  $ i=1,\dots,m,\ m:=N-r-s $. We introduce the following B-unitary matrix using the vectors defined so far:
	\begin{align}
		U=&(\tfrac{\boldsymbol{y}_1+\boldsymbol{z}_1}{2},\dots,\tfrac{\boldsymbol{y}_r+\boldsymbol{z}_r}{2},\boldsymbol{x}_1,\dots,\boldsymbol{x}_s,\boldsymbol{w}_1,\dots,\boldsymbol{w}_m,\nonumber \\
		&\tfrac{-\boldsymbol{y}_1+\boldsymbol{z}_1}{2},\dots,\tfrac{-\boldsymbol{y}_r+\boldsymbol{z}_r}{2},\tau \boldsymbol{x}_1^*,\dots,\tau \boldsymbol{x}_s^*,\tau \boldsymbol{w}_1^*,\dots,\tau \boldsymbol{w}_m^*).
	\end{align}
	Since the column vectors in this $ U $ form a $\sigma$-orthonormal basis, the following $ \sigma $-orthogonal relations hold: 
	\begin{align}
	\begin{split}
		&(\boldsymbol{x}_i,\boldsymbol{x}_j)_\sigma=-(\tau \boldsymbol{x}_i^*,\tau \boldsymbol{x}_j^*)_\sigma=\delta_{ij},\ (\boldsymbol{y}_i,\boldsymbol{z}_j)_\sigma=2\delta_{ij}, \\
		&(\boldsymbol{y}_i,\boldsymbol{y}_j)_\sigma=(\boldsymbol{z}_i,\boldsymbol{z}_j)_\sigma=(\boldsymbol{y}_i,\boldsymbol{x}_j)_\sigma=(\boldsymbol{y}_i,\tau \boldsymbol{x}_j^*)_\sigma=0, \\
		&(\boldsymbol{z}_i,\boldsymbol{x}_j)_\sigma=(\boldsymbol{z}_i,\tau \boldsymbol{x}_j^*)_\sigma=(\boldsymbol{x}_i,\tau \boldsymbol{x}_j^*)_\sigma=0,
	\end{split}\label{eq:sigmarelations}
	\end{align}
	where the relations for $ \boldsymbol{w}_i $'s are omitted. Using this $ U $, Colpa's standard form \cite{Colpa1986} for $ H_0 $ is given by
	\begin{align}
		U^{-1}H_0U=\begin{pmatrix} I_r &&&I_r && \\ &O_s&&&& \\ &&\Lambda&&& \\ -I_r&&&-I_r && \\ &&&&O_s& \\ &&&&&-\Lambda \end{pmatrix},
	\end{align}
	where $ \Lambda=\operatorname{diag}(\lambda_1,\dots,\lambda_m) $, and the spectral decomposition of  $ H_0 $ is given by
	\begin{align}
		H_0=\sum_{i=1}^m\lambda_i\boldsymbol{w}_i\boldsymbol{w}_i^\dagger\sigma+\sum_{i=1}^m\lambda_i\tau \boldsymbol{w}_i^*\boldsymbol{w}_i^T\tau\sigma+\sum_{i=1}^r\boldsymbol{y}_i\boldsymbol{y}_i^\dagger\sigma. \label{eq:Colpa01}
	\end{align}
	Note that this standard form is slightly different from our previous work \cite{DTMN}. In Ref.~\cite{DTMN}, if we use $ \tilde{\boldsymbol{y}}_i=\sqrt{\kappa_i}\boldsymbol{y}_i $ and $ \tilde{\boldsymbol{z}}_i=\boldsymbol{z}_i/\sqrt{\kappa_i} $ instead of $ \boldsymbol{y}_i $ and $ \boldsymbol{z}_i $, and if we omit tildes, then we obtain the expression in Eq.~(\ref{eq:Colpa01}) \footnote{For example, let us consider $ H_0=\left( \begin{smallmatrix}\kappa & \kappa \\ -\kappa & -\kappa \end{smallmatrix} \right) $. If we write $ \boldsymbol{y}=(1,-1)^T,\ \boldsymbol{z}=(1,1)^T, U=(\tfrac{\boldsymbol{y}+\boldsymbol{z}}{2},\tfrac{-\boldsymbol{y}+\boldsymbol{z}}{2})=I_2 $, then $ U^{-1}HU $ gives a standard form in Ref. \cite{DTMN}. Instead, if we use $ \tilde{\boldsymbol{y}}=(\sqrt{\kappa},-\sqrt{\kappa})^T,\ \tilde{\boldsymbol{z}}=(1/\sqrt{\kappa},1/\sqrt{\kappa})^T $ and define $ \tilde{U}=(\tfrac{\tilde{\boldsymbol{y}}+\tilde{\boldsymbol{z}}}{2},\tfrac{-\tilde{\boldsymbol{y}}+\tilde{\boldsymbol{z}}}{2}) $, then we indeed obtain  $ \tilde{U}^{-1}H_0\tilde{U}=\left( \begin{smallmatrix}1&1 \\ -1& -1 \end{smallmatrix} \right) $, corresponding to Eq.~(\ref{eq:Colpa01}). The B-unitarity of $ \tilde{U} $, i.e.,  $ \tilde{U}^{-1}=\sigma \tilde{U}^\dagger\sigma $ can be soon verified.}. %%%
The standard form in Ref.~\cite{DTMN} is unique under a different constraint, $ (\boldsymbol{y}_i,\boldsymbol{y}_j)_{\mathbb{C}}=2\delta_{ij} $, and this choice is convenient if the kinetic term is given by $ M_0=\sigma $. If the kinetic term is given by a more general matrix, however, this convention is not so convenient. \\
	\indent Next, let us calculate eigenvectors and eigenvalues of the matrix $ H_0+M_0k^2 $ for finite momentum $ k\ne 0 $ by perturbation theory. Let us expand eigenvectors and eigenvalues as  $ \boldsymbol{\xi}=\boldsymbol{\xi}_0+k\boldsymbol{\xi}_1+k^2\boldsymbol{\xi}_2+\dotsb $ and $ \epsilon=\epsilon_0+k\epsilon_1+k^2\epsilon_2+\dotsb $. Henceforth we are only interested in the cases where $ \xi_0 $ is an eigenvector of $ H_0 $ with zero eigenvalue. Thus we set $ \epsilon_0=0 $, and the perturbation equations up to  $ O(k^2) $ is given by
	\begin{align}
		H_0\boldsymbol{\xi}_1&=\epsilon_1\boldsymbol{\xi}_0 \label{eq:pertrb1st}\\
		M_0\boldsymbol{\xi}_0+H_0\boldsymbol{\xi}_2
&=\epsilon_2\boldsymbol{\xi}_0+\epsilon_1\boldsymbol{\xi}_1.\label{eq:pertrb2nd}
	\end{align}
	Since $ \boldsymbol{\xi}_0 $ is given by an eigenvector of $ H_0 $ with zero eigenvalue, and since the components of zeroth-order solutions in the higher-order terms $ \boldsymbol{\xi}_i $ with $ i\ge 1 $ can be always eliminated, we can set
	\begin{align}
		\boldsymbol{\xi}_0&=\sum_{j=1}^sa_j\boldsymbol{x}_j+\sum_{j=1}^sb_j\tau \boldsymbol{x}_j^*+\sum_{j=1}^rc_j\boldsymbol{y}_j, \\
		\boldsymbol{\xi}_l&=\sum_{j=1}^rd_j^{(l)}\boldsymbol{z}_j+\sum_{j=1}^{N-r-s}\alpha_j^{(l)}\boldsymbol{w}_j+\sum_{j=1}^{N-r-s}\beta_j^{(l)}\tau \boldsymbol{w}_j^*,\quad l\ge 1.
	\end{align}
	Form the first order equation (\ref{eq:pertrb1st}), we immediately have
	\begin{align}
		2d_i^{(1)}-\epsilon_1c_i=0,\quad \epsilon_1a_i=\epsilon_1b_i=0,\quad \alpha_i^{(1)}=\beta_i^{(1)}=0. \label{eq:firstorderperturb}
	\end{align}
	The next discussion differs depending on whether $ \epsilon_1 $ is zero or not.\\
	\indent We first consider the case  $ \epsilon_1\ne0 $. Then we obtain $ a_i=b_i=0 $ and $ d_i^{(1)}=\frac{1}{2}\epsilon_1c_i $. Thus, the eigenvector up to $ O(k^1) $ can be written as
	\begin{align}
		& \boldsymbol{\xi}_0=\sum_{j=1}^rc_j\boldsymbol{y}_j,\quad \boldsymbol{\xi}_1=\epsilon_1\sum_{j=1}^r\frac{c_j}{2}\boldsymbol{z}_j \\ \leftrightarrow& \quad \boldsymbol{\xi}=\sum_{j=1}^rc_j\left( \boldsymbol{y}_j+\frac{k\epsilon_1}{2}\boldsymbol{z}_j \right)+O(k^2).
	\end{align}
	Taking the  $ \sigma $-inner product between $ \boldsymbol{y}_i $ and the second-order equation (\ref{eq:pertrb2nd}), we obtain
	\begin{align}
		\sum_{j=1}^r(\boldsymbol{y}_i,M_0\boldsymbol{y}_j)_\sigma c_j=\epsilon_1^2 c_i.
	\end{align}
	If we define $ r\times r $ matrix $ Y $ whose $(i,j)$-component is given by $ Y_{ij}=(\boldsymbol{y}_i,M_0\boldsymbol{y}_j)_\sigma $, the above is the eigenvalue problem of $ Y $. 
	Since $ \sigma M_0 $ is assumed to be positive-definite, the matrix $ Y $ is positive-definite, real, and symmetric matrix. The fact that $Y$ is real can be checked as follows. If we write $ \boldsymbol{y}_j=(\boldsymbol{\phi}_j,-\boldsymbol{\phi}_j^*)^T $, then
	\begin{align}
		(\boldsymbol{y}_i,M_0\boldsymbol{y}_j)_\sigma=2\operatorname{Re}\left( \boldsymbol{\phi}_i^\dagger M\boldsymbol{\phi}_j-\boldsymbol{\phi}_i^\dagger L \boldsymbol{\phi}_j^* \right),
	\end{align}
	which is obviously real. Therefore, there exist a real orthogonal matrix $ R $ such that $ R^{-1}YR $ becomes diagonal, and the eigenvalues are all real and positive. If we introduce a new basis by $ \tilde{\boldsymbol{y}}_i=\sum_i \boldsymbol{y}_j R_{ji} $ and $ \tilde{\boldsymbol{z}}_i=\sum_j\boldsymbol{z}_jR_{ji} $, and write the eigenvalues as $ 2\kappa_1,\dots, 2\kappa_r (>0) $, 
	\begin{align}
		(\tilde{\boldsymbol{y}}_i,M_0\tilde{\boldsymbol{y}}_j)_\sigma=2\kappa_i\delta_{ij},\quad 2\kappa_1,\dots,2\kappa_r>0. \label{eq:diagonaltypei}
	\end{align}
	Thus, the first order eigenvalue is given by $ \epsilon_1=\pm\sqrt{2\kappa_i} $, giving the linear dispersion  $ \epsilon=\pm\sqrt{2\kappa_i}k+O(k^2) $, and the eigenvector is given by  $ \tilde{\boldsymbol{y}}_i\pm k\sqrt{\frac{\kappa_i}{2}}\tilde{\boldsymbol{z}}_i+O(k^2) $. Here we note that the tilde-added vectors, $ \tilde{\boldsymbol{y}}_i $'s and $  \tilde{\boldsymbol{z}}_i $'s also satisfy the same $ \sigma $-orthogonal relations in Eq.~(\ref{eq:sigmarelations}). \\
	\indent Next, let us consider the case $ \epsilon_1=0 $. From Eq.~(\ref{eq:firstorderperturb}), we have $ d_i^{(1)}=\alpha_i^{(1)}=\beta_i^{(1)}=0 $ and hence $ \boldsymbol{\xi}_1=0 $. Thus the perturbation equation begins from the second-order, given by
	\begin{align}
		M_0\boldsymbol{\xi}_0+H_0\boldsymbol{\xi}_2=\epsilon_2\boldsymbol{\xi}_0. \label{eq:2ndorderprtrb}
	\end{align}
	We first introduce the following vectors  $ \tilde{\boldsymbol{x}}_i $'s by the Gram-Schmidt-like process:
	\begin{align}
		\tilde{\boldsymbol{x}}_i=\boldsymbol{x}_i-\sum_{j=1}^r\frac{(\tilde{\boldsymbol{y}}_j,M_0\boldsymbol{x}_i)_\sigma}{2\kappa_j}\tilde{\boldsymbol{y}}_j.
	\end{align}
	The corresponding  $ \tau\tilde{\boldsymbol{x}}_i^* $ can be written in the same form:
	\begin{align}
		\tau\tilde{\boldsymbol{x}}_i^*=\tau \boldsymbol{x}_i^*-\sum_{j=1}^r\frac{(\tilde{\boldsymbol{y}}_j,M_0\tau \boldsymbol{x}_i^*)_\sigma}{2\kappa_j}\tilde{\boldsymbol{y}}_j.
	\end{align}
	This can be shown as follows. Since $ M_0 $ and $ \sigma $ are B-hermitian, $ \tau M_0^* \tau=-M_0 $ and $ \tau\sigma\tau=-\sigma $ hold. Noting them and the relation $ \tilde{\boldsymbol{y}}_j=-\tau \tilde{\boldsymbol{y}}_j^* $, we have
	\begin{align}
		&(\tilde{\boldsymbol{y}}_j,M_0 \boldsymbol{x}_i)_\sigma^*=\tilde{\boldsymbol{y}}_j^T\sigma M_0^* \boldsymbol{x}_i^*=(\tilde{\boldsymbol{y}}_j^T\tau)(\tau\sigma\tau)(\tau M_0^*\tau)\tau \boldsymbol{x}_i^*\nonumber \\
		&=-\tilde{\boldsymbol{y}}_j^\dagger \sigma M_0\tau \boldsymbol{x}_i^*=-(\tilde{\boldsymbol{y}}_j,M_0\tau \boldsymbol{x}_i^*)_\sigma.
	\end{align}
	The new basis  $ \tilde{\boldsymbol{x}}_i,\ \tau\tilde{\boldsymbol{x}}_i^* $ do not change the $ \sigma $-orthogonal relations in Eq.~(\ref{eq:sigmarelations}), and further satisfy the following:
	\begin{align}
		(\tilde{\boldsymbol{x}}_i,M_0\tilde{\boldsymbol{y}}_j)_\sigma=(\tau\tilde{\boldsymbol{x}}_i^*,M_0\tilde{\boldsymbol{y}}_j)_\sigma=0.
	\end{align}
	Since $ M_0 $ is B-hermitian, the relation $ (M_0\tilde{\boldsymbol{x}}_i,\tilde{\boldsymbol{y}}_j)_\sigma=(M_0\tau\tilde{\boldsymbol{x}}_i^*,\tilde{\boldsymbol{y}}_j)_\sigma=0 $ also holds. Then, let us redefine the starting zeroth order eigenvector $ \boldsymbol{\xi}_0 $ as
	\begin{align}
		\boldsymbol{\xi}_0=\sum_{j=1}^sa_j\tilde{\boldsymbol{x}}_j+\sum_{j=1}^sb_j\tau \tilde{\boldsymbol{x}}_j^*+\sum_{j=1}^rc_j\tilde{\boldsymbol{y}}_j.
	\end{align}
	This redefinition does not change the result of the first-order perturbation calculations in Eq.~(\ref{eq:firstorderperturb}).  Then, taking the $ \sigma $-inner product between the second-order equation (\ref{eq:2ndorderprtrb}) and $ \tilde{\boldsymbol{y}}_j $, and using Eq.~(\ref{eq:diagonaltypei}), we obtain
	\begin{align}
		c_j=0,\quad j=1,\dots,r.
	\end{align}
	Next, taking the  $ \sigma $-inner products between Eq.~(\ref{eq:2ndorderprtrb}) and $ \tilde{\boldsymbol{x}}_i $ or $ \tau \tilde{\boldsymbol{x}}_i^* $, we obtain 
	\begin{align}
		\sum_{j=1}^s(\tilde{\boldsymbol{x}}_i,M_0\tilde{\boldsymbol{x}}_j)_\sigma a_j+\sum_{j=1}^s(\tilde{\boldsymbol{x}}_i,M_0\tau\tilde{\boldsymbol{x}}_j^*)_\sigma b_j=\epsilon_2a_i, \\
		-\sum_{j=1}^s(\tau\tilde{\boldsymbol{x}}_i^*,M_0\tilde{\boldsymbol{x}}_j)_\sigma a_j-\sum_{j=1}^s(\tau\tilde{\boldsymbol{x}}_i^*,M_0\tau\tilde{\boldsymbol{x}}_j^*)_\sigma b_j=\epsilon_2b_i.
	\end{align}
	Now, let $ X $ and $  \Xi $ be $ s\times s $ matrices whose $ (i,j) $-component is given by  $ X_{ij}=(\tilde{\boldsymbol{x}}_i,M_0\tilde{\boldsymbol{x}}_j)_\sigma $ and $  \Xi_{ij}=(\tilde{\boldsymbol{x}}_i,M_0\tau\tilde{\boldsymbol{x}}_j^*)_\sigma $, respectively. Then, the above equations are interpreted as the eigenvalues problem of the following B-hermitian matrix $ Z $:
	\begin{align}
		Z=\begin{pmatrix}X & \Xi \\ -\Xi^* & -X^* \end{pmatrix}.
	\end{align}
	Due to the assumption that $ \sigma M_0 $ is positive-definite,  $ \sigma Z $ is also positive-definite. Thus, from the theorem of Ref.~\cite{Colpa1978} (or from Theorem 3.4 of Ref.~\cite{DTMN}), there exists a B-unitary matrix $ U $ such that 
	\begin{align}
		&U^{-1}ZU=\operatorname{diag}(\mu_1^{-1},\dots,\mu_s^{-1},-\mu_1^{-1},\dots,-\mu_s^{-1}),\nonumber \\
		& \mu_1,\dots,\mu_s>0.
	\end{align}
	If we write new basis vectors diagonalizing $ Z $ as  $ \tilde{\tilde{\boldsymbol{x}}}_i,\ \tau \tilde{\tilde{\boldsymbol{x}}}_i^* $, the dispersion relation of type-II mode arising from $ \tilde{\tilde{\boldsymbol{x}}}_i $ is given by $ \epsilon=\mu_i^{-1}k^2+O(k^4) $, and that from $ \tau\tilde{\tilde{\boldsymbol{x}}}_i^* $ is given by $ \epsilon=-\mu_i^{-1}k^2+O(k^4) $. We thus obtain type-II dispersion relations. \\
	\indent Finally we add a remark. If we rewrite the tilde-added vectors $ \tilde{\boldsymbol{y}}_j,\ \tilde{\tilde{\boldsymbol{x}}}_j $ with tildeless notations as $ \boldsymbol{y}_j,\ \boldsymbol{x}_j $, then they satisfy the following $ \sigma $-orthogonal relations:
	\begin{align}
		(\boldsymbol{x}_i,M_0\boldsymbol{x}_j)_\sigma=(\tau \boldsymbol{x}_i^*,M_0\tau \boldsymbol{x}_j^*)_\sigma=\frac{1}{\mu_i}\delta_{ij},\ (\boldsymbol{x}_i,M_0\tau \boldsymbol{x}_j^*)_\sigma=0,\\
		(\boldsymbol{y}_i,M_0\boldsymbol{y}_j)_\sigma=2\kappa_i\delta_{ij},\quad (\boldsymbol{y}_i,M_0\boldsymbol{x}_j)_\sigma=(\boldsymbol{y}_i,M_0\tau \boldsymbol{x}_j^*)_\sigma=0.
	\end{align}
	If we set $ M_0=\sigma $ in these relations, it becomes a revisit of the $ \sigma $-orthogonal relations given in 
Subsec.~4.1 of Ref.~\cite{DTMN}. The derivation shown here is also applicable to the case $ M_0=\sigma $. The derivation here means that the perturbative calculations and derivations of type-I and type-II dispersion relations do not need the block-diagonalization of the WB matrix, if we appropriately solve the perturbative equation for degenerate zero eigenvalues. However, in the special case $ M_0=\sigma $, as was shown in Subsec.~2.3 of Ref.~\cite{DTMN}, the choice of the basis such that the WB matrix becomes block-diagonal makes perturbative calculations a little easier.

%\bibliography{QNGcomplexONmodel}% Produces the bibliography via BibTeX.

\end{document}